\documentclass[12pt]{article}
\usepackage{times}
\usepackage{geometry}
\usepackage{graphicx}
\usepackage[utf8]{inputenc}
\usepackage{enumitem,amssymb}
\usepackage{ragged2e}
\newlist{thematic}{itemize}{8}
\setlist[thematic]{label=$\square$}
\usepackage{pifont}
\usepackage{wrapfig}

\begin{document}
\raggedright
\huge
Astro2020 Science White Paper \linebreak

Understanding Exoplanet Atmospheres with UV Observations I: NUV and Blue/Optical\linebreak

\normalsize
\noindent \textbf{Thematic Areas:} \hspace*{60pt} $\boxtimes$ Planetary Systems \hspace*{10pt} $\boxtimes$ Star and Planet Formation \hspace*{20pt}\linebreak
$\square$ Formation and Evolution of Compact Objects \hspace*{31pt} $\square$ Cosmology and Fundamental Physics \linebreak
  $\square$  Stars and Stellar Evolution \hspace*{1pt} $\square$ Resolved Stellar Populations and their Environments \hspace*{40pt} \linebreak
  $\square$    Galaxy Evolution   \hspace*{45pt} $\square$             Multi-Messenger Astronomy and Astrophysics \hspace*{65pt} \linebreak
  
\textbf{Principal Author:}

Name: Jessie Christiansen  
 \linebreak                                             
Institution: Caltech/IPAC-NExScI
 \linebreak
Email: jessie.christiansen@caltech.edu
 \linebreak
Phone: (626) 720-9649
 \linebreak
 
\textbf{Co-authors:} 
  \linebreak
  Thomas Barclay (GSFC/U. Maryland), Luca Fossati (OeAW), Kevin France (U. Colorado), Eric Lopez (GSFC), Jason Rowe (Bishop's University), Joshua Schlieder (GSFC), Hannah Wakeford (STScI), Allison Youngblood (GSFC)
  \linebreak
  
\textbf{Co-signers:}
  \linebreak
  Karan Molaverdikhani (MPIA/Heidelberg Univ.), Diana Dragomir (MIT/UNM), Vincent Bourrier (Observatory of Geneva, U. Geneva), Kerri Cahoy (MIT), Chuanfei Dong (Princeton University), Shawn Domagal-Goldman (NASA GSFC), Giada Arney (NASA GSFC), David Ardila (Jet Propulsion Laboratory), Evgenya Shkolnik (Arizona State University), Erika Hamden (University of Arizona), Roxana Lupu (BAER Institute / NASA Ames), Kevin Hardegree-Ullman (Caltech/IPAC-NExScI), Douglas Caldwell (SETI Institute), Knicole D. Col\'on (NASA GSFC), Courtney D. Dressing (University of California, Berkeley), Jasmina Blecic (New York University Abu Dhabi), Henry Ngo (National Research Council Canada), Daniel Angerhausen (Bern Unisersity), Alain Lecavelier des Etangs (Institut d'astrophysique de Paris, CNRS, France), Kathleen Mandt (Johns Hopkins University Applied Physics Laboratory), Seth Redfield (Wesleyan University), Jonathan Fortney (USCS), Dawn Gelino (Caltech/IPAC-NExScI), David R. Ciardi (Caltech/IPAC-NExScI), Jake D. Turner (Cornell University), Kyle A. Pearson (University of Arizona) \linebreak

\textbf{Abstract:}
Much of the focus of exoplanet atmosphere analysis in the coming decade will be at infrared wavelengths, with the planned launches of the James Webb Space Telescope (JWST) and the Wide-Field Infrared Survey Telescope (WFIRST). However, without being placed in the context of broader wavelength coverage, especially in the optical and ultraviolet, infrared observations produce an incomplete picture of exoplanet atmospheres. Scattering information encoded in blue optical and near-UV observations can help determine whether muted spectral features observed in the infrared are due to a hazy/cloudy atmosphere, or a clear atmosphere with a higher mean molecular weight. UV observations can identify atmospheric escape and mass loss from exoplanet atmospheres, providing a greater understanding of the atmospheric evolution of exoplanets, along with composition information from above the cloud deck. In this white paper we focus on the science case for exoplanet observations in the near-UV; an accompanying white paper led by Eric Lopez will focus on the science case in the far-UV.

\pagebreak

\section{Exoplanet Atmospheres: Current Limitations}
\label{sec:limits}

Our sample of exoplanets orbiting bright host stars will expand rapidly as the NASA {\it TESS} mission realises its full potential. The promise of these exoplanets is in the characterization of their composition, both bulk and atmospheric. Prioritizing planets for which atmospheric observations will yield the greatest insight will be crucial for maximizing science returns.\newline 

\begin{wrapfigure}{r}{0.5\linewidth}
\centering
\includegraphics[width=0.45\textwidth]{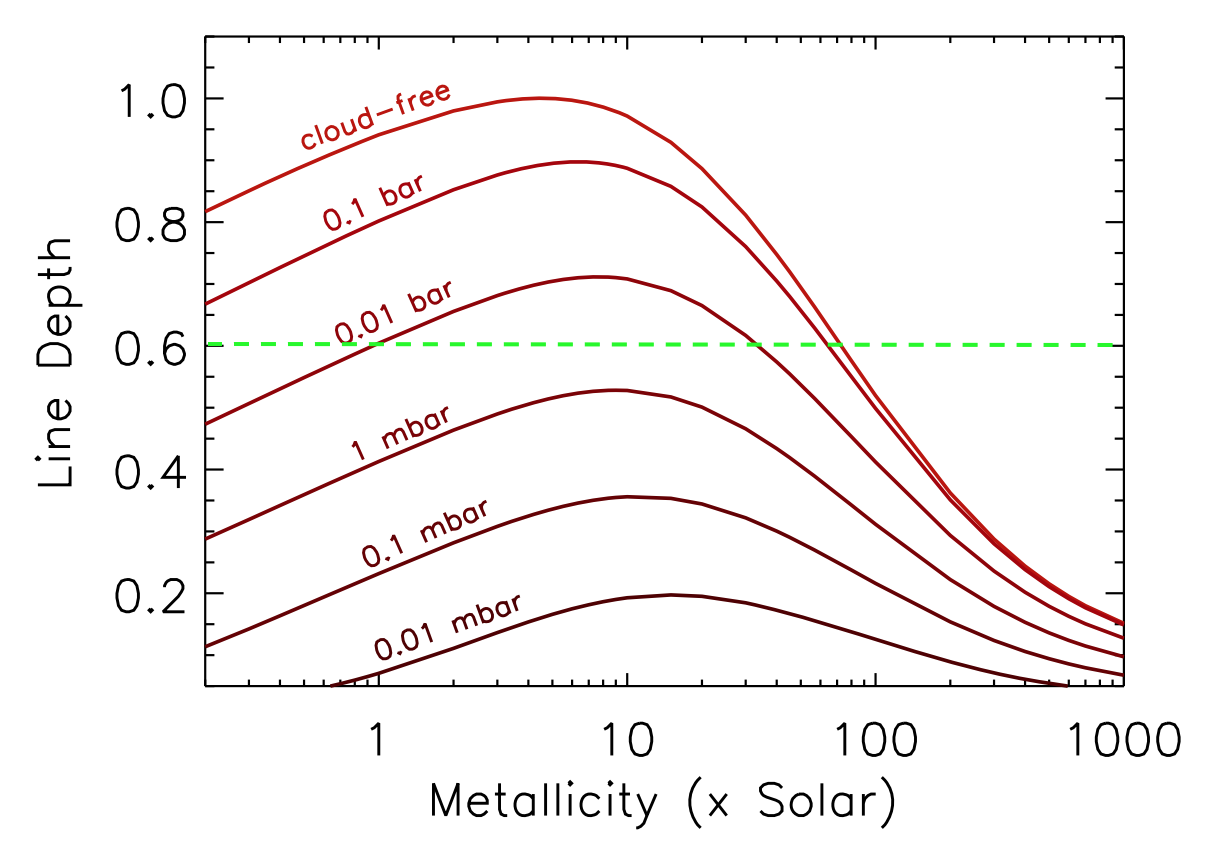}
\caption{Reproduced from Fig. 7 of Kempton et al. (2017), showing spectral line depths as a function of metallicity. The pressure lines indicate the height of the cloud deck.}
\label{fig:kempton2017b}
\end{wrapfigure}

Transmission spectroscopy---measuring the wavelength-dependent absorbing cross section of the planet during transit---has been one of the most productive paths of investigation. 
One discovery is that clouds and hazes are prevalent in all types of exoplanet atmospheres (Wakeford et al. 2019; Fu et al. 2017; Iyer et al 2016). In particular, hazes may become more significant for planets cooler than 850~K (Crossfield \& Kreidberg 2017; Morley et al. 2015). Both can give rise to muted or largely featureless spectra, especially in the infrared. However, similar spectra can also be caused by atmospheres with a high mean molecular weight (Madhusudhan \& Redfield 2015). Sing et al. (2016) find in their large study of 10 hot Jupiter planets that the amplitude of the water features vary significantly, ``ranging from features that are very pronounced (as in WASP-19b) to those that are significantly smaller than expected (HD 209458b) or even absent (WASP-31b).'' This diversity holds for warm Neptunes (e.g. Fig. 1 of Crossfield \& Kreidberg 2017), and super-Earths (Southworth et al. 2017). 
Figure \ref{fig:kempton2017b} (reproduced from Fig 7 of Kempton et al. 2017) shows the suppression of spectral line depth as a function of metallicity ($\propto$mean molecular weight), for cloud decks ranging from clear to 0.01~mbar. The final line depth is degenerate with both metallicity and the cloud deck location, for example the same spectral line depth could be caused by either solar metallicity atmosphere with clouds at 0.01 bar, or a clear atmosphere with 100$\times$ solar metallicity. This degeneracy has hindered further understanding of these muted spectra. 
Looking ahead to the launch of the James Webb Space Telescope (JWST), being able to (a) predict which planets are most likely to have identifiable spectral features, and (b) interpret the very high quality infrared transmission spectra that will be obtained will be crucial for maximising the return of this flagship mission.\newline

One advantage that we can exploit is that observations at different wavelengths probe different parts of an exoplanet atmosphere. The near-UV (NUV) and blue optical are often overlooked wavelength regions for exoplanet studies as they require high signal-to-noise ratio (SNR) observations and therefore bright and close-by exoplanets. Yet, this wavelength region contains unique constraints on the atmospheric properties of the planet, including the scattering properties of cloud and haze particles (e.g., Wakeford \& Sing 2015), and evidence for atmospheric escape (e.g., Bourrier, et al. 2018). We explore these science cases in this white paper. \newline 

\vspace{-0.5in}
\section{NUV/Blue optical transmission spectroscopy}
\label{sec:uvobs}

\begin{wrapfigure}{r}{0.55\linewidth}
\Centering
\includegraphics[width=0.55\textwidth]{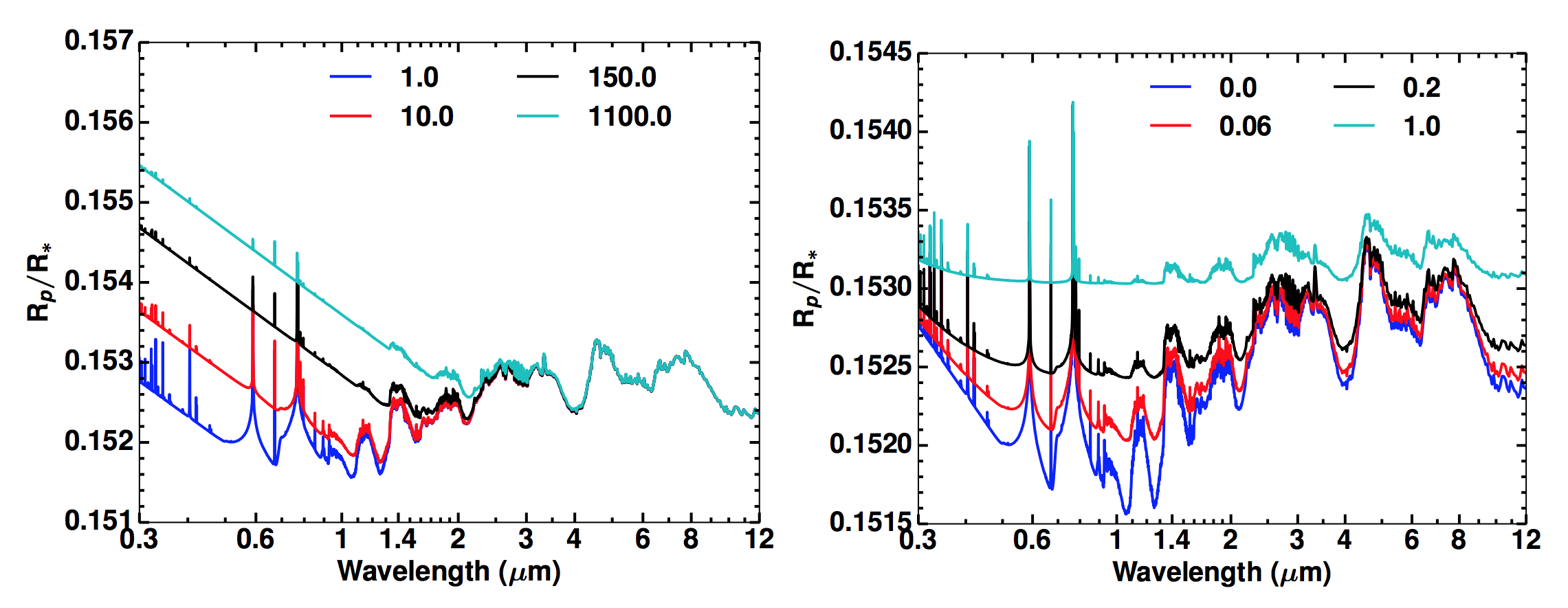}
\caption{Reproduced from Figures 17a and 17b from Goyal et al. (2018), showing the model HD 189733~b transmission spectra with varying haze (left panel) and cloud (right panel) enhancement factors.}
\label{fig:goyal2018}
\end{wrapfigure}

The ability of observations at NUV/blue optical wavelengths to help determine the atmospheric composition lies in the properties of the clouds and hazes expected to be present. Due to our current lack of observational constraints on the types of aerosol particles possible, clouds (formed by reversible condensation chemistry) are typically treated in models as large, primarily scattering particles with grey opacity, obscuring spectral features across all wavelengths (e.g. Goyal et al. 2018). On the other hand, hazes (formed by irreversible 

\begin{wrapfigure}{r}{0.55\linewidth}
\centering
\includegraphics[width=0.55\textwidth]{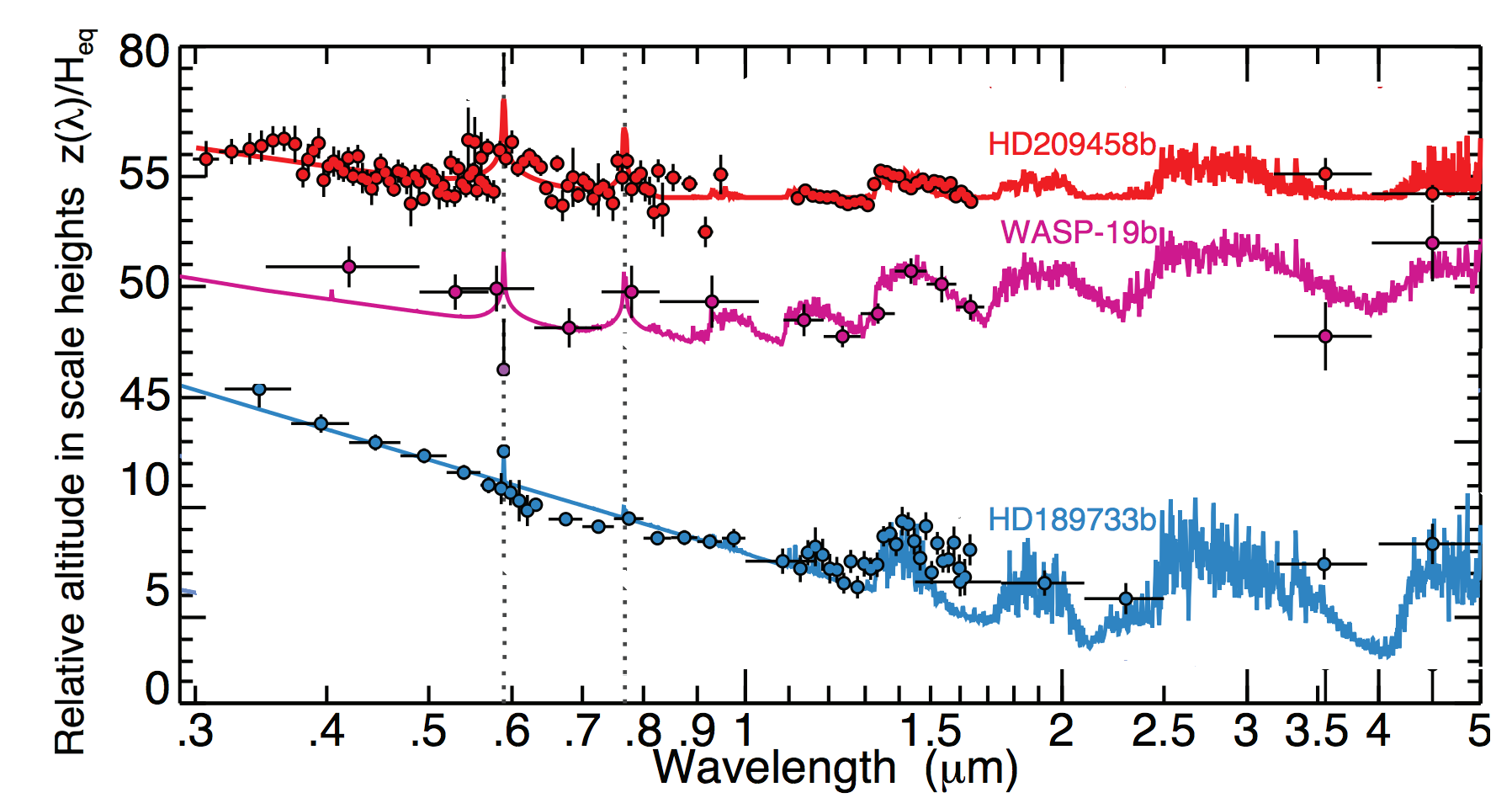}
\caption{Modified from Fig. 1 of Sing et al. (2016)}
\label{fig:sing2016}
\end{wrapfigure}

photochemistry) are treated as small scattering particles with a Rayleigh-like scattering profile ($\propto \lambda^{-4}$). The impact of varying the levels of clouds and hazes in the atmosphere can be seen in Figure \ref{fig:goyal2018}, where the largest differences are seen at optical/NUV wavelengths.\newline 

This effect can clearly be seen in the transmission spectra of the best observationally-constrained and understood exoplanets, HD~209458~b and HD~189733~b, shown in Figure \ref{fig:sing2016}, where HD 189733~b shows one of the strongest enhanced Rayleigh scattering signatures, seen in the striking difference in slope shorter than 1$\mu m$. On the other hand, exoplanets with few observations in the blue optical/NUV (e.g. WASP-19b, also shown in Figure \ref{fig:sing2016}) are challenging to interpret.\newline

As a result, several studies have found that an effective metric for differentiating atmosphere types is by anchoring the NIR transmission spectra with optical/NUV measurements (Sing et al. 2016; Wakeford et al. 2018). Wakeford et al. (2018) find that including the optical/NUV observations in their analysis of WASP-39~b allows them to greatly reduce the wide range of models allowed by the NIR data alone, which would otherwise allow for strong hazes, completely cloud- or haze-free models, and models with both near solar and highly super-solar metallicity. Sing et al. (2016) define a metric, $\Delta Z_{\rm UB-LM}$, which measures the difference in the strength of the spectral features at 0.3--0.57~$\mu m$ (capturing the scattering) and those at 3--5~$\mu m$ (capturing the molecular absorption); they encourage measurement of this and several other indices to identify clear atmospheres in advance of future costly observations. In their tentative detection of water in the atmosphere of the super-Earth GJ 1132~b, Southworth et al. (2017) also call for future observations at wavelengths $<$0.5$\mu m$ to help differentiate between competing model spectra.





\section{Atmospheric Escape}

Atmospheric escape is a key factor shaping the exoplanet population, and planet formation theories rely on the demographics of this population. The upper atmospheres of planets, their interactions with the host stars, and escape processes can be best studied observationally at UV wavelengths, because at longer wavelengths the optical depth of the escaping material is typically too low. UV spectroscopy has revealed the presence of an hydrodynamically escaping atmosphere for a handful of close-in giant exoplanets. Most of these observations have been conducted in the FUV (see the accompanying white paper by Eric Lopez), focusing particularly on Ly\,$\alpha$, but NUV observations have been successfully used to study upper atmospheres and escape (Fossati et al. 2010; Haswell et al. 2012; Vidal-Madjar et al. 2013; Salz et al. 2019).\newline 

Although not covering hydrogen lines, observations at NUV wavelengths present several advantages over those in the FUV. Importantly, the NUV photosphere is not as strongly variable as the FUV chromosphere, where clear detections of low SNR can be difficult (Loyd \& France  2014). The continuum emission of solar-like stars rises in the FUV-NUV transition region, thus in the NUV the flux occulted by the planet is more smoothly distributed across the stellar disc, hence the transit depth can be more directly related to the fraction of occulted area (Haswell 2010). The NUV spectral region includes several strong lines of metals present and absorbing in the escaping atmospheres of giant planets (e.g., MgI, MgII, FeII, MnII) and a continuum probing scattering by aerosols. 
Furthermore, in the NUV, because of the higher stellar flux, it is possible to study more distant systems (up to several hundred parsecs; e.g., Fossati et al. 2010), than is possible in the FUV. This is due to the fact that the NUV does not suffer from strong interstellar attenuation from atomic resonance lines (except Mg II) like the species used for FUV transit measurements (Ly $\alpha$, C II, O I,  etc)

\section{Host star UV emission}


Characterization of the absolute flux and spectral properties of the host star's UV emission is necessary to properly interpret species like O$_2$, O$_3$, CH$_4$, and CO$_2$ detected in exoplanet atmospheres (France et al. 2016). Across the UV, these molecules have large photo-absorption cross sections that vary with wavelength by orders of magnitude. For host stars whose UV spectral energy distributions are vastly different than the Sun's (e.g., M dwarfs), detectable quantities of byproducts of life on Earth like O$_2$ and O$_3$ can build up through photo-chemistry alone on CO$_2$-rich planets (Domagal-Goldman et al. 2014; Tian et al. 2014; Harman et al. 2015). In addition, orders of magnitude flux increases on short timescales due to stellar flares have been regularly observed (e.g., Hawley \& Pettersen 1991; Hawley 2002; and many more). Loyd et al.~(2017,2018) indicate that emission from flares may actually dominate the UV emission of M dwarf stars, so the impact of UV flares on exoplanet photochemistry could be substantial.\newline

To fully characterize a host star's UV emission, direct observations are recommended. Stellar synthetic spectral models that self-consistently treat the chromosphere, transition, and corona (the photosphere does not dominate the UV flux of cool stars) do not exist, and current semi-empirical models are not easily set up to run for a large number of stars (Fontenla et al. 2016; Peacock et al. 2019). The incident NUV flux is an important input to exoplanet photochemical models and there are a swathe of important metal emission lines across the near and far UV, which are important both for characterizing stellar activity and for understanding atmospheric escape. In particular, the Mg II\,h\&k lines at $\approx$280~nm and the Ca II K Line at $\approx$393~nm have been shown to be a good proxies for the critical key Lyman $\alpha$ feature in the FUV (Muscles II \& IV, Youngblood et al. 2016, 2017). For further information on observations of escaping atmospheres and the Lyman $\alpha$ line, see the accompanying FUV white paper by Eric Lopez. 





\section{Current ultraviolet capabilities}
\label{sec:current}

Due to the presence of the ozone layer in our own atmosphere, observations in the blue optical/NUV are increasingly hindered towards shorter wavelengths, and virtually impossible shorter than $\sim0.3\mu m$. The majority of the current and planned UV capability for exoplanet characterization is therefore in space.

\subsection{Hubble Space Telescope}
The Hubble Space Telescope (HST) has been instrumental in revealing the UV-optical-IR transmission properties for exoplanet atmospheres and has already revealed a number of trends; 1) clouds and hazes are prevalent in all exoplanet atmospheres, yet it is difficult to predict which exoplanet atmospheres they will effect the most; 2) exoplanets exhibit extended upper atmospheres with detection of atmospheric escape lending clues as to the evolution of planetary atmospheres. The NUV is vital to further address these two science topics and HST has the unique capability of measuring these features across a wide wavelength range ($\sim$150-500\,nm). The exoplanet atmospheres observed so far show Rayleigh-like scattering in the U and B band wavelength ranges (Sing, et al., 2016), primarily due to scattering caused by small particles (e.g., Lecavelier des Etangs, 2008; Wakeford, et al., 2017). As described above, transmission spectra probe the planetary atmosphere as a function of altitude via the wavelength regime; looking toward the UV provides access to lower pressures in their atmospheres at higher altitudes where scattering by cloud particles becomes dominant in the absorption spectrum. The Space Telescope Imaging Spectrograph (STIS) E230M and G430L gratings 
have been the workhorse for NUV/blue optical investigations for nearly a decade (see for example the spectra in Figure \ref{fig:sing2016}).\newline



\section{Future ultraviolet prospects}
\label{sec:future}

\subsection{CASTOR}


The Canadian Space Agency is currently planning for the Cosmological Advanced Survey Telescope for Optical and uv Research (CASTOR; PI Patrick C\^ot\'e) mission, a 1~m telescope covering the 0.135--0.55~$\mu m$ blue optical to FUV wavelength range with imaging and spectroscopic capabilities. The telescope will operate close to the diffraction limit, with a proposed field of view that is about two orders of magnitude larger than that of the Hubble Space Telescope. The proposed 5-year mission will include legacy surveys and Guest Observer components.  Two exoplanet programs have been proposed to use the simultaneous UV, u, g imaging capabilities to survey 50 bright (g $<$ 10) transiting exoplanets to measure precise transit depths and phase-curves.  The observation will explore the diversity of transit and phase curve colours in the NUV due to scattering and NUV molecular absorption.


\subsection{The small-sat opportunity}


With the increasing capability and decreasing cost of small satellite missions, there is opportunity to explore these NUV/blue optical science cases in a focused way. As noted by Fleming et al. (2018): "Small satellite missions enable the study of transient phenomena over extended time periods in a manner not feasible for large, multi-purpose space observatories such as HST."\newline

One such satellite selected for launch by NASA is the Colorado Ultraviolet Transit Experiment (CUTE; PI Kevin France). The telescope has an effective collecting area of $\sim$30~cm$^2$, and will cover 255--333~nm at a resolving power of R$\sim$3000. In the short mission duration (nominally seven months, extendable to two years), the team plans to collect transmission spectroscopy of at least 12 short-period (1--5 days) exoplanets (24--30 in an extended mission). Repeat visits ($\sim$10 per system) will allow the team to compile large numbers of transits of each exoplanet, and subsequently to explore the exoplanet transmission and host star variability.\newline

Another future option that is well motivated by the science cases described here is an explorers-class mission. A small aperture (20--30~cm) telescope that covers a broad wavelength range (250--900~nm) and carries multi-color photometry or low-resolution spectroscopy capable of transmission spectroscopy would be a cost efficient way to make progress on assembling a large catalog of well-characterized exoplanet systems. Such capabilities, even on non-exoplanet specific missions, have great utility. This is exemplified by the recent detection of an NUV exoplanet transit by the UVOT instrument on the Neil Gehrels {\it Swift} Observatory (Salz et al. 2019)

\section{Conclusions}

In this white paper we have highlighted the importance of characterizing exoplanets and their host stars in the often over-looked blue optical/NUV wavelengths. An accompanying white paper by Eric Lopez discusses the value of observing in the FUV. Crucial interpretation of observations at other wavelengths relies on information that can only be readily obtained at these shorter wavelengths. We end with the note that care is needed when combining information from multiple wavelengths---observations taken at very different times will be complicated by stellar variability. 

\pagebreak
\textbf{References}\newline

Bourrier, V., Lecavelier des Etangs, A., Ehrenreich, D., et al.\ 2018, A\&A, 620, A147\\
Crossfield, I.~J.~M., \& Kreidberg, L.\ 2017, AJ, 154, 261\\
Domagal-Goldman, S.~D., Segura, A., Claire, M.~W., et al.\ 2014, ApJ, 792, 90\\
Fleming, B.~T., France, K., Nell, N., et al.\ 2018, JATIS, 4, 14004\\
Fontenla, J.~M., Linsky, J.~L., Witbrod, J., et al.\ 2016, ApJ, 830, 154\\
Fossati, L., Haswell, C., Froning, C., et al. 2010, ApJL, 714, L222\\
France, K., Loyd, R.~O.~P., Youngblood, A., et al.\ 2016, AhP, 820, 89\\
Fu, G., Deming, D., Knutson, H., et al.\ 2017, ApJ, 847, L22\\
Goyal, J.~M., Mayne, N., Sing, D.~K., et al.\ 2018, MNRAS, 474, 5158\\
Harman, C.~E., Schwieterman, E.~W., Schottelkotte, J.~C., et al.\ 2015, ApJ, 812, 137\\
Haswell, C. 2010, Transiting exoplanets, Cambridge University Press, ISBN: 9780521139380\\
Haswell, C., Fossati, L., Ayres, T., et al. 2012, ApJ, 760, 79\\
Hawley, S.~L., Covey, K.~R., Knapp, G.~R., et al.\ 2002, AJ, 123, 3409\\
Hawley, S.~L., \& Pettersen, B.~R.\ 1991, ApJ, 378, 725\\
Iyer, A.~R., Swain, M.~R., Zellem, R.~T., et al.\ 2016, ApJ, 823, 109\\
Kempton, E.~M.-R., Lupu, R., Owusu-Asare, A., et al.\ 2017, PASP, 129, 44402\\
Lecavelier Des Etangs, A., Vidal-Madjar, A., D{\'e}sert, J.-M., et al.\ 2008, A\&A, 485, 865\\
Loyd, R.~O.~P., France, K., Youngblood, A., et al.\ 2018, ApJ, 867, 71\\
Loyd, R.~O.~P., Koskinen, T.~T., France, K., et al.\ 2017, ApJ, 834, L17\\
Loyd, R.~O.~P., \& France, K.\ 2014, ApJS, 211, 9\\
Madhusudhan, N., \& Redfield, S.\ 2015, International Journal of Astrobiology, 14, 177\\
Morley, C.~V., Fortney, J.~J., Marley, M.~S., et al.\ 2015, ApJ, 815, 110\\
Peacock, S., Barman, T., Shkolnik, E.~L., et al.\ 2019, ApJ, 871, 235\\
Salz, M., Schneider, P. C., Fossati, L., et al. 2019, A\&A, 623, A57\\
Sing, D.~K., Fortney, J.~J., Nikolov, N., et al.\ 2016, Nature, 529, 59\\
Southworth, J., Mancini, L., Madhusudhan, N., et al.\ 2017, AJ, 153, 191\\
Tian, F., France, K., Linsky, J.~L., et al.\ 2014, EAPS Letters, 385, 22\\
Vidal-Madjar, A., Huitson, C., Bourrier, V., et al. 2013, A\&A, 560, 54\\
Wakeford, H.~R., Wilson, T.~J., Stevenson, K.~B., et al.\ 2019, RNAAS, 3, 7\\
Wakeford, H.~R., Sing, D.~K., Deming, D., et al.\ 2018, AJ, 155, 29\\
Wakeford, H.~R., Sing, D.~K., Kataria, T., et al.\ 2017, Science, 356, 628\\
Wakeford, H.~R., \& Sing, D.~K.\ 2015, A\&A, 573, A122\\
Youngblood, A., France, K., Loyd, R.~O.~P., et al.\ 2017, ApJ, 843, 31\\
Youngblood, A., France, K., Loyd, R.~O.~P., et al.\ 2016, ApJ, 824, 101\\

\end{document}